\documentclass[aps,prb,superscriptaddress,preprint,showpacs,amsmath,amssymb,floatfix,11pt]{revtex4-1}

\usepackage{graphicx}
\usepackage{bm}
\usepackage[breaklinks=true,colorlinks,citecolor=blue,linkcolor=blue,urlcolor=blue]{hyperref}
\usepackage{color}
\usepackage{amsmath}

\begin{document}

\title{Gate-tunable spatial modulation of localized plasmon resonances}

\author{Andrea Arcangeli}
\affiliation{NEST, Scuola Normale Superiore and Istituto Nanoscienze-CNR, Piazza san Silvestro 12, I-56127 Pisa, Italy}

\author{Francesco Rossella}
\email{francesco.rossella@sns.it}
\affiliation{NEST, Scuola Normale Superiore and Istituto Nanoscienze-CNR, Piazza san Silvestro 12, I-56127 Pisa, Italy}

\author{Andrea Tomadin}
\affiliation{NEST, Scuola Normale Superiore and Istituto Nanoscienze-CNR, Piazza san Silvestro 12, I-56127 Pisa, Italy}

\author{Jihua Xu}
\affiliation{NEST, Scuola Normale Superiore and Istituto Nanoscienze-CNR, Piazza san Silvestro 12, I-56127 Pisa, Italy}

\author{Daniele Ercolani}
\affiliation{NEST, Scuola Normale Superiore and Istituto Nanoscienze-CNR, Piazza san Silvestro 12, I-56127 Pisa, Italy}

\author{Lucia Sorba}
\affiliation{NEST, Scuola Normale Superiore and Istituto Nanoscienze-CNR, Piazza san Silvestro 12, I-56127 Pisa, Italy}

\author{Fabio Beltram}
\affiliation{NEST, Scuola Normale Superiore and Istituto Nanoscienze-CNR, Piazza san Silvestro 12, I-56127 Pisa, Italy}

\author{Alessandro Tredicucci}
\affiliation{NEST, Scuola Normale Superiore and Istituto Nanoscienze-CNR, Piazza san Silvestro 12, I-56127 Pisa, Italy}
\affiliation{Dipartimento di Fisica, Universit\`a di Pisa, Largo Pontecorvo 3, I-56127 Pisa, Italy}

\author{Marco Polini}
\affiliation{Istituto Italiano di Tecnologia, Graphene Labs, Via Morego 30, I-16163 Genova,~Italy}
\affiliation{NEST, Scuola Normale Superiore, Piazza dei Cavalieri 7, I-56126 Pisa,~Italy}

\author{Stefano Roddaro}
\affiliation{NEST, Scuola Normale Superiore and Istituto Nanoscienze-CNR,  
Piazza san Silvestro 12, I-56127 Pisa, Italy}

\maketitle

{\bf Nanoplasmonics~\cite{Maier07,stockman11} exploits the coupling between light and collective electron density oscillations (plasmons~\cite{Pines_and_Nozieres,Giuliani_and_Vignale}) to bypass the stringent limits imposed by diffraction. This coupling enables confinement of light to sub-wavelength volumes and is usually exploited in nanostructured metals. Substantial efforts are being made at the current frontier of the field to employ electron systems in semiconducting and semimetallic materials since these add the exciting possibility of realizing electrically tunable and/or active nanoplasmonic devices~\cite{Kim2012,Chen2008,Chen2009,Fei2012,Chen2012a}. Here we demonstrate that a suitable design of the doping profile in a semiconductor nanowire (NW) can be used to tailor the plasmonic response and induce localization effects akin to those observed in metal nanoparticles~\cite{Klar1998}. Moreover, by field-effect carrier modulation, we demonstrate that these localized plasmon resonances can be spatially displaced along the nanostructure body, thereby paving the way for the implementation of {\em spatially} tunable plasmonic circuits.}

The excitation of localized plasmon resonances (LPRs) leads to a strong local enhancement of the electromagnetic fields, as a consequence of the collective oscillation of free charge carriers.
This found important applications in sensing~\cite{Mayer2011}, microscopy~\cite{Rothenhausler1988}, Raman spectroscopy~\cite{Kneipp1997}, and photovoltaics~\cite{Mubeen2014,Atwater2010}.  Traditionally, LPRs were observed and investigated in {\em metallic} nanoparticles (NPs), but recent experiments are extending this possibility to doped semiconductor NPs~\cite{Kramer2015, Luther2011}. One of the key motivations for the exploration of this route is the perspective of obtaining materials with lower free carrier density $n$ and thus lower plasma frequency $\omega_{\rm p}$. In doped semiconductors, this yielded the observation of LPRs in nanoparticles from the mid- to the far-infrared region of the spectrum. A further important drive pushing for this paradigmatic shift to semiconductors is the perspective of electrical tuning of the plasmonic response. The investigation of hybrid electronic/plasmonic systems is indeed at its infancy and holds the promise of novel nano-optoelectronic chips embedding multiple functionalities~\cite{Sorger2012}. A remarkable example is provided by the recent progress in graphene plasmonics~\cite{Grigorenko2012}, where the high room-temperature mobility of samples encapsulated in hexagonal boron nitride crystals allows gate-controlled tuning of long-lived and highly-confined two-dimensional plasmons~\cite{Woessner2014}. 

In our experiment, we exploit state-of-the-art growth of self-assembled semiconductor NWs~\cite{Lucia}, which offer the unique flexibility to engineer local properties of the semiconductor. 
In particular, we use steep axial doping profiles in InAs NWs to localize the system's plasmonic response and control the {\it spatial} position of the NW LPR by field-effect.

\begin{figure*}[h!]
\includegraphics[width=6in]{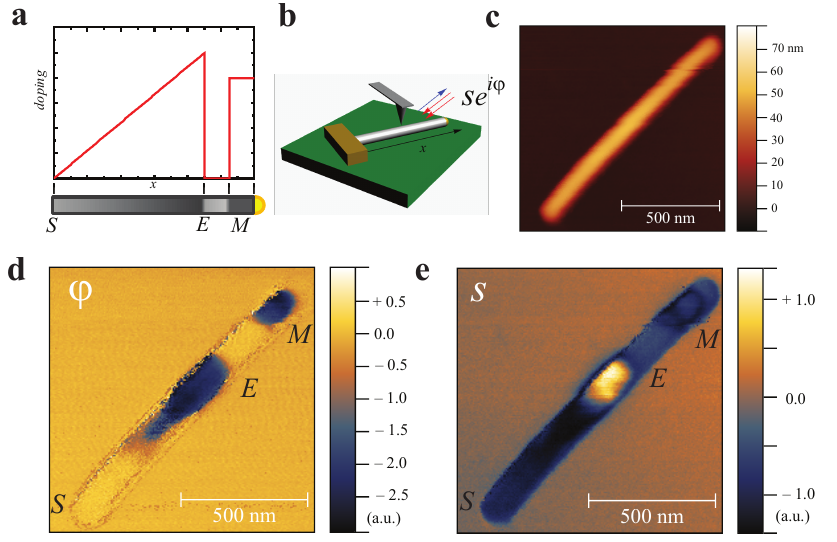}
\caption{\label{fig:one}
{\bf Local plasmon resonance in a steep doping profile.} {\bf a}, By a suitable process described in the main text, we create a linearly varying carrier density profile $n(x)$ along the growth direction in an indium arsenide nanowire. The induced density ranges from a nominally undoped value $n\approx1\times10^{16}\,{\rm cm^{-3}}$ (in the region labeled by the letter $S$) to a maximum doping $n\approx5\times10^{18}\,{\rm cm^{-3}}$ (in the region labeled by the letter $E$). Intermediate-doping segments were introduced in the growth sequence and used as markers ($M$) for the s-SNOM maps. {\bf b}, NWs were deposited on a SiO$_2$/n-Si substrates and a $\lambda = 10.5\,{\rm \mu m}$ laser beam (red arrows) is focused on an s-SNOM tip oscillating at $250\,{\rm kHz}$. The amplitude $s$ and phase $\varphi$ of the reflected beam (blue arrow) are detected using an interferometric pseudo-heterodyne technique, demodulated at the fourth harmonic of the tip tapping frequency and used to reveal the local dielectric response of the NW (see Methods). {\bf c}, AFM topography map of an isolated NW, acquired in parallel to the SNOM signal. {\bf d}, Fourth harmonic s-SNOM amplitude ($s_4$) map: the plasmon mode is highlighted by a strong modulation of the scattered field amplitude in proximity of the region where the laser matches the local plasma frequency. {\bf e}, Fourth harmonic s-SNOM phase ($\varphi_4$) map: a marked phase dip in the scattered light is observed in correspondence with the plasma frequency. The doping marker is visible as a region of modified scattering (both in amplitude and phase) and provides a reference to identify the graded doping region. 
}
\end{figure*}

A prototypical example of the investigated NWs is shown in Fig.~\ref{fig:one}. A set of $60-80\,{\rm nm}$-diameter InAs wires were grown by metal-seeded chemical beam epitaxy and included a $\approx 1\,{\rm \mu m}$-long steep graded-doping region. In nominally undoped portions of the NW, the free-carrier concentration $n$ was of the order of $10^{16}~{\rm cm}^{-3}$ while a carrier density up to $n\approx5\times10^{18}\,{\rm cm}^{-3}$ is estimated for the highest doping-density regions, as determined by transport measurements in standard field-effect transistors grown in equivalent conditions~\cite{Viti2012}. The plasmonic response of the NWs was investigated by scattering-type scanning near-field optical microscopy (s-SNOM) in the mid-infrared spectral range~\cite{Stiegler2010a,Hillenbrand2000}. We used a metallic tip with a radius of $\sim 20~{\rm nm}$ operating in tapping mode at $250\,{\rm kHz}$ to probe the near-field optical response of the sample under a mid-infrared CO$_{2}$ laser ($\lambda = 10.5~{\rm \mu m}$). A pseudo-heterodyne interferometric detection was exploited to obtain the amplitude ($s$) and phase ($\varphi$) of the scattered optical beam (Fig.~\ref{fig:one}b), which strongly depends on the local dielectric response of the NW. In our experiment, the backscattered signal was demodulated at the fourth harmonic of the probe tapping frequency. This procedure minimizes far-field contributions (see Methods) and yields the amplitude component $s_4$, with phase $\varphi_4$. 

Atomic force microscope (AFM) (Fig.~\ref{fig:one}c) and s-SNOM (Figs.~\ref{fig:one}d,e) maps of isolated NWs on SiO$_2$/n-Si were acquired simultaneously. The optical response of the sample reveals the localized nature of the plasmonic resonance in the graded-doping nanostructure. The maps in Figs.~\ref{fig:one}c,d show the amplitude $s_4$ and phase $\varphi_4$ of the scattered field as a function of the SNOM position on the sample, respectively. Enhanced scattering only occurs in a small region of the NW where the local plasma frequency $\omega_{\rm p}(x)$ matches the illumination frequency $\omega$. We see that the plasmon resonance extends for $\approx 200\,{\rm nm}$ and no indication of plasmon propagation is observed. 
Our NWs support LPRs for a variety of reasons. First, contrary to strictly 2D systems like graphene, our NW hosts a quasi-three-dimensional electron system with bulk-like plasmons. These have a dispersion relation that is nearly flat~\cite{Pines_and_Nozieres,Giuliani_and_Vignale} in the relevant region of wave vectors where tip-plasmon coupling occurs. This implies a nearly zero plasmon wave vector and group velocity. Second, due to the presence of donors, which have been implanted in the body of the NW to engineer the steep doping density profile, the lifetime of our bulk-like plasmons is very short. Within a simple Drude model for the NW local conductivity $\sigma(\omega)$, we find a plasmon lifetime $\tau$ on the order of $10~{\rm fs}$. Third, the highly inhomogeneous carrier density distribution $n(x)$ along the NW growth direction in our NWs enables highly-selective spatial coupling between the SNOM tip and the plasmon. This is because the plasmon frequency is a function of the local carrier density (see Methods), $\omega_{\rm p}(n) = \sqrt{4\pi n e^2 /(\epsilon_{\infty} m_{\rm b})}$, and therefore, for a given illumination frequency $\omega$, energy conservation implies tip-plasmon coupling only occurs in the neighborhood of the position $x_{\rm max}$ at which $\omega = \omega_{\rm p}(n(x_{\rm max}))$.

\begin{figure*}

\includegraphics[width=5in]{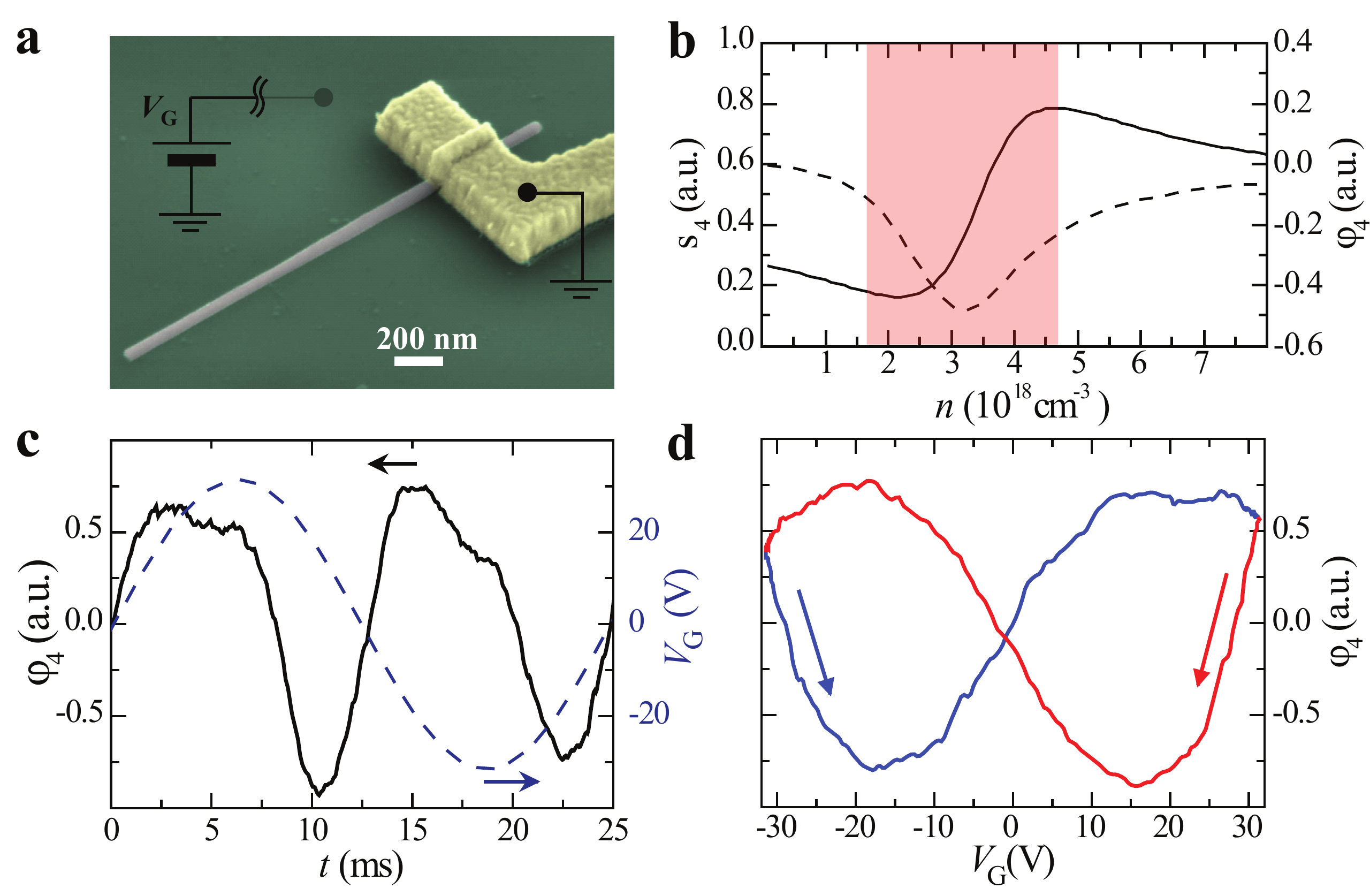}
\caption{\label{fig:mod}
{\bf Field-effect modulation of the plasma resonance.} {\bf a}, NWs were deposited on a SiO$_2$/n-Si substrate and contacted by a Ni/Au electrode, which allows using the n-Si substrate as a backgate. {\bf b}, Calculated fourth harmonic amplitude $s_4$ and phase $\varphi_4$ modulation of the s-SNOM signal, as predicted by a point-dipole tip-sample interaction model (see Methods). The curves are obtained by assuming a Drude-like dielectric constant for the InAs NW. Using a laser excitation at $\lambda=10.5\,{\rm \mu m}$, a strong modulation is expected for a carrier density $n\approx3\times10^{18}\,{\rm cm^{-3}}$. The charge density interval covered by the resonance is highlighted by the red shaded region. {\bf c}, Time evolution of the $\varphi_4$ signal (black line) caused by a sinusoidal modulation of the gate voltage (blue dashed line). During the measurement, the s-SNOM probe is kept on top of the resonant region on the NW at zero backgate voltage. A full resonance modulation is achieved by sweeping the gate voltage from $-30\,{\rm V}$ to $+30\,{\rm V}$ at $40\,{\rm Hz}$. {\bf d}, Phase modulation as a function of the gate voltage $V_G$. A strong hysteresis is observed by sweeping the gate voltage from low to high values (blue curve) and back (red curve).}
\end{figure*}

In order to attain the field-effect control of the LPR in the nanostructures we fabricated devices equipped with an electrical reference contact and a gate allowing carrier-density manipulation (Fig.~\ref{fig:mod}a). NWs were dropcasted on top of a SiO$_2$/n-Si substrate as in the case of Fig.~\ref{fig:one}. Using electron beam lithography, we defined a Ni/Au electrode at the very end of each NW in order to use the n-Si substrate as a backgate and minimize electrostatic screening~\cite{Pitanti2012}. Our device structure allows the application of up to $\pm 80\,{\rm V}$ gate voltage sweeps in few milliseconds. Based on simple electrostatic modelling~\cite{Wunnicke}, the capacitance $C_{\rm g}$ between the backgate and a NW with a diameter in the range $60$-$80~{\rm nm}$ is estimated to be $\approx 35$-$38~{\rm aF/\mu m}$, corresponding to a carrier density modulation in the range $7.6$-$12.4\times 10^{16}~{\rm cm}^{-3}/{\rm V}$. The impact of field-effect on the NW plasmon response was measured by positioning the s-SNOM tip on top of the NW resonant region (at $\lambda \approx  10.5~{\rm \mu m}$) and by measuring the tip-induced light scattering as a function of gate voltage $V_G$. The expected carrier-density dependence of the amplitude and phase of the scattered field are shown in Fig.~\ref{fig:mod}b. The resonance is expected to occur at $n\approx 3\times10^{18}~{\rm cm}^{-3}$ and a carrier density modulation $\delta n$ of at least $3\times10^{18}~{\rm cm}^{-3}$ is necessary to map the full plasmonic resonance (see shaded area in Fig.~\ref{fig:mod}b). The actual measured NW response is shown in Figs.~\ref{fig:mod}c,d. The experimental configuration was particularly stable and unaffected by sample mechanical drift with respect to the s-SNOM tip. Multiple gate scans could thus be acquired and averaged to achieve good signal-to-noise ratios. In Fig.~\ref{fig:mod}c we report the time evolution of the phase $\varphi$ of the scattered wave, when the gate voltage $V_{\rm G}$ is subject to a peak-to-peak oscillation of $60~{\rm V}$ at $40~{\rm Hz}$. A  strong modulation of the optical response is observed. In Fig.~\ref{fig:mod}d the phase of the scattered light is plotted against $V_{\rm G}$. The marked hysteresis is a consequence of charge trapping at the NW/dielectric interface~\cite{Astromskas2010,Roddaro2013}.

\begin{figure*}
\includegraphics[width=6in]{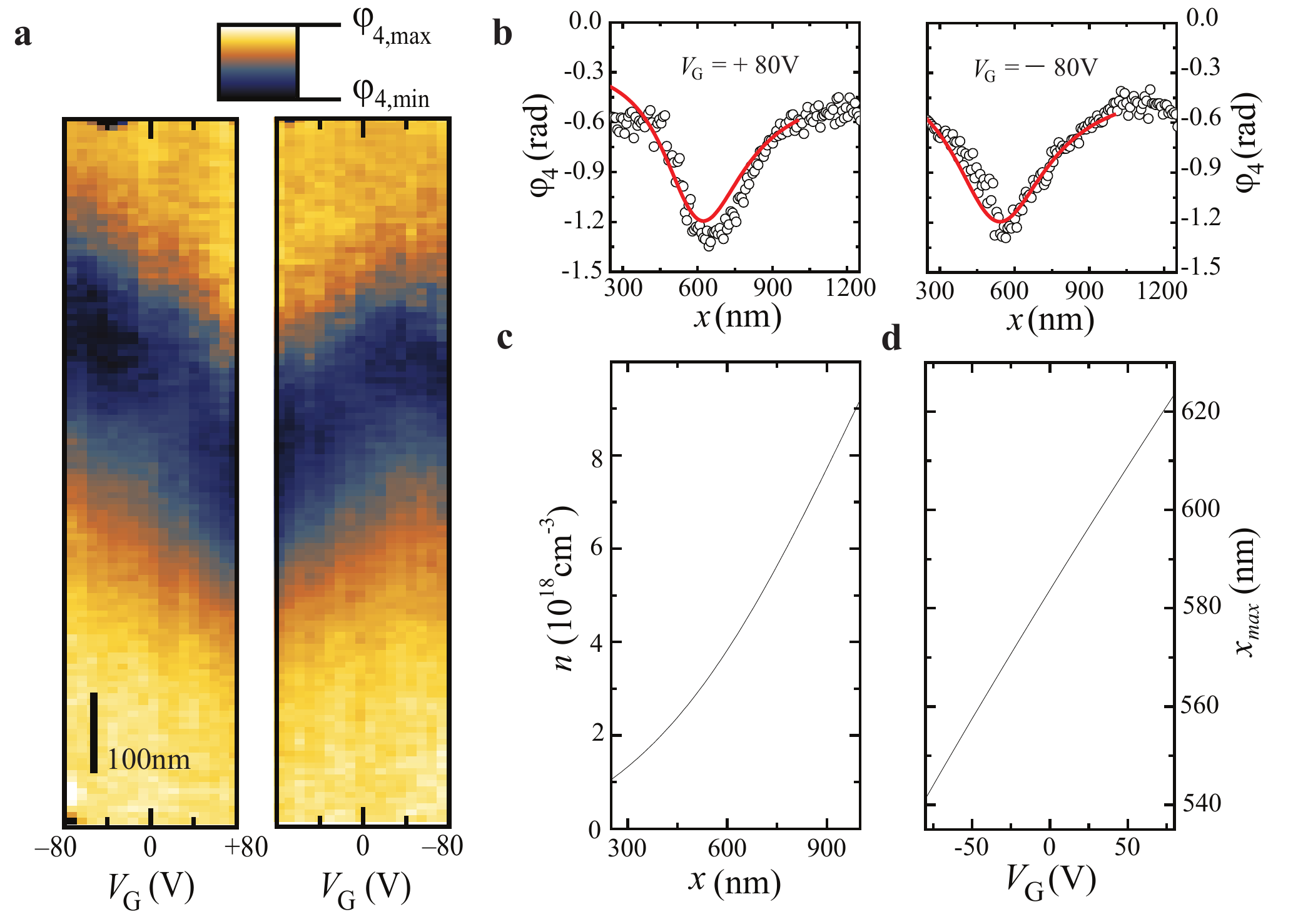}
\caption{\label{fig:scan}
{\bf Spatial shift of the local plasmon resonance.} 
{\bf a}, Color plot of the evolution of the s-SNOM phase profiles measured along the NW growth direction, as a function of the gate voltage $V_{\rm G}$. The displacement of the LPR along the NW axis is observed as the gate voltage $V_{\rm G}$ is controlled from $-80,{\rm V}$ to $+80\,{\rm V}$ and from $+80\,{\rm V}$ to $-80\,{\rm V}$.  {\bf b}, Fit to the experimental data for the extremal values of the gate voltage. The phase modulation is well reproduced by the point-dipole interaction model and the full field-effect evolution of the NW response is reproduced with a single fitting procedure. {\bf c}, Estimate of the charge density profile along the NW axis. {\bf d}, Estimate of the displacement of the LPR position along the NW axis as a function of the $V_{\rm G}$: the curve is obtained using the values for the gate capacity and the doping profile extracted by fitting the experimental data in the plot of panel a. 
}
\end{figure*}

The modulation of light scattering displayed in Fig.~\ref{fig:mod} corresponds to a physical {\it displacement} of the LPR along the NW axis. Crucially, this is enabled by the fact that the gate electrode, being capacitively coupled to the NW, shifts the carrier density profile $n(x)$ by a constant value $\delta n$ which is proportional to the gate voltage $V_{\rm G}$. In turn, since the NW is designed to feature a linear spatial dependence of $n(x)$ versus the longitudinal coordinate $x$, the energy conservation condition $\omega = \omega_{\rm p}(n(x))$ is expected to occur at a gate-shifted position $x_{\rm max} + \delta x_{\rm max}$ where $\delta x_{\rm max}=-\delta n\cdot(dn/dx)^{-1}\propto V_{\rm G}$. To confirm this, we measured near-field infrared profiles $\varphi_4(x)$ along the NW growth axis while varying the gate voltage from $-80\,{\rm V}$ to $+80\,{\rm V}$ and then back to $-80\,{\rm V}$ (Fig.~\ref{fig:scan}a). In order to rule out possible spurious effects, e.g. stemming from small piezoelectric deformations of the SiO$_2$ layer~\cite{Lazovski2012}, we monitored the position of the metallic reference electrode and/or of the doping marker regions at each scan. Experimental data clearly indicate that the plasma resonance is shifted along the NW axis by about $100\,{\rm nm}$ when the gate voltage is swept from $-80\,{\rm V}$ to $+80\,{\rm V}$ and viceversa. This corresponds to a shift of the resonance by more than its half width at half maximum. We note that, compared to the measurements shown in Fig.~\ref{fig:mod}, a larger gate voltage swing is needed to achieve a modulation covering the full resonance. This is not surprising since traps on the NW surface and in the gate dielectric are known to cause non-negligible screening effects that are particularly evident at room temperature and for slow gating sweeps~\cite{Astromskas2010,Roddaro2013}. Indeed, the time required to acquire each s-SNOM scan is of the order of several minutes, while in the case of the measurements reported in Fig.~\ref{fig:mod}, a full gate sweep lasted only few milliseconds. It is important to note that this is not an intrinsic limitation of the devices but is rather dictated by the experimental need to perform slow s-SNOM scans.

The experimental data are well described by a point-dipole interaction model \cite{Cvitkovic2007}, from which the complete charge-density profile along the NW axis can be derived. The s-SNOM signal was compared with the expression for the fourth-order harmonic of the scattering field induced by the tip-sample interaction (see Methods). We fit the expression for $\varphi_{4}(x)$ by modeling the carrier density profile $n(x)$ with a constant term $n_{0}$, a uniform contribution proportional to the gate voltage $V_{\rm G}$, a linearly-increasing term, and a first-order non-linear correction:
\begin{equation}\label{eq:density}
n(x) =  n_{0} - V_{\rm G} C / e + a_{0} x + a_{1} \sin{(\pi x / b)}~.
\end{equation}
Here, $C$ is a capacitance per unit volume and $b=1\,{\rm \mu m}$ is the length of the graded doping region. We remark that a single set of parameters was fitted against all curves at different $V_{\rm G}$. The result of the fit for the two largest values of $V_{\rm G}$ is shown in Fig.~\ref{fig:scan}b, while the deduced density profile $n(x)$ is shown in Fig.~\ref{fig:scan}c. Probably owing to the partial shielding by the grounded electrode, the carrier profile $n(x)$ is slightly sublinear for $x \lesssim 500~{\rm nm}$. Starting from the best fit and using the expression for $\varphi_4$, we obtained the position $x_{\rm max}$ of the resonance (Fig.~\ref{fig:scan}d). We estimated an LPR displacement of $0.6~{\rm nm/V}$ driven by the field effect. As mentioned before, in this experimental configuration the LPR shift is hindered by screening effects driven by traps in the dielectric and on the NW surface, which can become significant particularly at slow gating sweeps. This is also confirmed by the low apparent value of $C$ extracted from the fit of the experimental data, when compared with the expected value $C_{\rm g}$ (see Methods).

In conclusion, we demonstrated that we can localize and spatially control by field effect the position of plasmon resonances within InAs nanowires by embedding a suitably engineered doping profile. Our novel nanoplasmonic architecture combines the strong local light confinement typically observed for metallic nanoparticles with the wide tunability typical of semiconductors. The achievement of electrically-tunable plasmon systems is of great relevance to applications: in the specific case of tunable LPRs, it can pave the way to innovative devices concepts such as, for instance, recently proposed scanning plasmon sensors~\cite{Nishi2015}. In general, present results on local doping modulation indicate a promising direction for the design and implementation of a new class of devices with electrical spatial control of plasmonic modes. 

\vspace{1cm} 
\noindent {\bf Methods}

Electron-doped InAs nanowire (NW) growths were performed by Au-assisted chemical beam epitaxy (CBE) in a Riber Compact 21 system. The system employs pressure control in the metalorganic (MO) lines to determine precursor fluxes during sample growth. A calibrated orifice at the injector insures the proportionality between line pressure and precursor flow through the injector. The precursors involved in the NW growth are tri-methylindium (TMIn), tertiarybutylarsine (TBAs), and ditertiarybutyl selenide (DtBSe) as selenium source for $n$-type doping. 
A nominally 0.5 nm thick Au film was first deposited on (111)B InAs wafers by thermal evaporation. Before the growth was initiated, the sample was heated at 470 $\pm$ 10 $^circ$C under TBAs flow for 20 min in order to dewet the Au film into nanoparticles and to remove the surface oxide from the InAs substrate. InAs stems were grown at a temperature of 380 $\pm$ 10 $^circ$C for 30 min with MO line pressures of 0.3 and 0.7 Torr for TMIn and TBAs, respectively. The growth temperature was increased to 400 $\pm$ 10 $^circ$C for the growth of the graded-doped InAs NWs. The steep doping profile along the axial direction of the NWs was obtained by tuning the partial pressure of di-tert-butyl selenide (DtBSe) from 0 to 1 Torr during the growth process.

Near-field measurements were performed by a scattering-type scanning near-field optical microscope (NeaSNOM from Neaspec GmbH).  It is based on an atomic force microscope (AFM) operating in tapping mode at a frequency of about 250 kHz. Pt-coated AFM tips with a radius of R $\simeq$ 20 nm were chosen as near-field probes.  The tip was illuminated with a commercial (accesslaser.com) grating-tunable carbon dioxide (CO$_{2}$) laser emitting in a TEM$_{00}$ transverse mode (M$^2$=1). Radiation (about 1 mW) was focused on the tip by means of a 2 cm focal length parabolic mirror. The emission of the laser can be tuned between 9.5 and 10.7 $\mu$m. The system is equipped with a pseudo-heterodyne interferometric detection module in order to detect simultaneously the phase and amplitude of the scattered signal. The optical signal was acquired simultaneously with topographic AFM maps. The signals acquired in this work were the scattering amplitude and phase, demodulated at the fourth harmonic of the tapping frequency.

In the fit of the experimental data we assumed a Drude form for the dielectric function of the carriers in the NW: 
$\epsilon_{\rm s}(\omega) = \epsilon_{\infty} [ 1 - \omega_{\rm pl}^{2} / (\omega^{2} - i \omega / \tau) ]$.
We introduced the plasma frequency $\omega_{\rm pl} = [4 \pi n e^{2} / (\epsilon_{\infty} m_{\rm b})]^{1/2}$
where $m_{\rm b} = 0.023~m_{\rm e}$ is the carriers' band mass, $m_{\rm e}$ is the electron mass in vacuum, and $\epsilon_{\infty} = 15.15$ is the high-frequency dielectric constant of InAs.
We took the Drude scattering time $\tau = 10.0~{\rm fs}$, which yields a Drude mobility $\mu = e \tau / m_{\rm b} \simeq 0.75 \times  10^{3}~{\rm cm}^{2}/({\rm V}{\rm s})$ in line with the range of values reported in the literature~\cite{Viti2012}. The expression for the fourth-order scattering contribution of the s-SNOM signal is

\begin{equation}\label{eq:phase}
s_{4} e^{i \varphi_{4}} = a \frac{3}{8192} \alpha^{2} \beta (1+\beta) \left ( 80 + 68 \alpha \beta + 5 \alpha^{2} \beta^{2} \right ) \left ( 1 - \frac{\alpha\beta}{4} \right )^{-5}~,
\end{equation}

\noindent where $a$ is a real constant which relates the near- and far-field of the radiating point dipole, $\alpha \equiv (\epsilon_{\rm t} - 1)/(\epsilon_{\rm t} + 2)$, $\beta \equiv [\epsilon_{\rm s}(\omega) - 1] / [\epsilon_{\rm s}(\omega) + 1]$, with $\epsilon_{\rm t}$ and $\epsilon_{\rm s}(\omega)$ the dielectric constant of the tip material and the dielectric function in the NW, respectively. The fit yields the following best values for the parameters in Eq.~(\ref{eq:density}): $a_{0} \simeq 0.91 \times 10^{23}~{\rm cm}^{-4}$, $a_{1} \simeq -177 \times 10^{16}~{\rm cm}^{-3}$, and $C \simeq 0.87 \times 10^{-15}~{\rm F} \, {\rm \mu m}^{-3}$. The latter in turn provides a value for the gate capacitance  $C_{\rm g}$ between the backgate and the NW in the range $2.5$-$4.4~{\rm aF/\mu m}$. This value is different from what expected from simple electrostatics. Well-known analytical models~\cite{Wunnicke} in fact estimate the back-gate capacitance per unit length as $C_g = 2\pi\epsilon_{\rm SiO_2}/arccosh[(d_{\rm SiO_2}+r)/r]$, where $\epsilon_{\rm SiO_2} = 3.9\epsilon_0$ is the oxide permittivity, $\epsilon_0$ is vacuum permittivity, $d_{\rm SiO_2}$ is thickness of the SiO$_2$ layer, and $r$ is the radius of the NW. Using $d_{rm SiO_2} = 285\,{\rm nm}$ and $r = 30-40\,{\rm nm}$, we estimate $C_g = 35-38\,{\rm aF/\mu m}$. Correspondingly, the expected carrier density modulation $dn/dV_G = C_g/\pi r^2$ is in the range
$7.6-12.4 \times 10^{16}\,{\rm cm^{−3}/V}$. The mismatch with the calculated values of $35-38\,{\rm aF/\mu m}$ is ascribed to the different impact of screening by charge traps at the NW/dielectric interface, depending on the gate sweep speed. The gate sweep speeds
during the measurements can be estimated to be (i) up to
about $\pm 8\,{\rm V/ms}$ in point modulation data of Figure 2, (ii) about $\pm0.1\,{\rm V/s}$ in the gate-modulated scans of Figure 3, and (iii) about $3\,{\rm V/s}$ during the preparatory step before the scans of Figure 3 (sweep from $V_G = 0$ to $+80\,{\rm V}$ in about $30\,{\rm s}$). The shift of the LPR was observed on five different devices. For three of these, the field effect modulation of the plasmon resonance was demonstrated both by looking at the SNOM response in a single position (Figure 2) and by performing a spatial s-SNOM map the resonance line as a function of $V_G$ (Figure 3). Depending on the average $V_G$ sweep speed, the resonance shift in the s-SNOM maps was estimated to fall in the $0.6-1.6\,{\rm nm/V}$ range.

\acknowledgments
We thank I. Torre, M.S. Vitiello, F. De Angelis and R.M. Macfarlane for useful discussions. This work was supported by the EC under the Graphene Flagship program (contract no.~CNECT-ICT-604391), the MIUR through the programs ``FIRB - Futuro in Ricerca 2013'' - Project ``UltraNano'' (Grant No.~RBFR13NEA4) and ``Progetti Premiali 2012'' - Project ``ABNANOTECH'', and by the ERC advanced grant SoulMan (G.A. 321122). Free software was used (www.gnu.org, www.python.org).

\end{document}